\documentclass[twocolumn,superscriptaddress,showpacs,eqsecnum]{revtex4}
\usepackage{times,xspace}
\usepackage{amsbsy,amssymb,amsmath,bm}
\usepackage{graphicx,color,epsfig,rotate}
\usepackage{fancyhdr}

\def\bbbc{{\mathchoice {\setbox0=\hbox{$\displaystyle\rm C$}\hbox{\hbox
to0pt{\kern0.4\wd0\vrule height0.9\ht0\hss}\box0}}
{\setbox0=\hbox{$\textstyle\rm C$}\hbox{\hbox
to0pt{\kern0.4\wd0\vrule height0.9\ht0\hss}\box0}}
{\setbox0=\hbox{$\scriptstyle\rm C$}\hbox{\hbox
to0pt{\kern0.4\wd0\vrule height0.9\ht0\hss}\box0}}
{\setbox0=\hbox{$\scriptscriptstyle\rm C$}\hbox{\hbox
to0pt{\kern0.4\wd0\vrule height0.9\ht0\hss}\box0}}}}

\newcommand{\bS}{\boldsymbol S}

\newcommand{\br}{\boldsymbol r}

 \newcommand{\hate}{\hat{\text{\rm e}}}

\newcommand{\ignore}[1]{}
\newcommand{\mComment}[1]{}
\newcommand{\gComment}[1]{}
\newcommand{\jComment}[1]{}
\newcommand{\rComment}[1]{}
\newcommand{\lComment}[1]{}

\renewcommand{\mComment}[1]{\textcolor{blue}{Manny: #1}}
\renewcommand{\gComment}[1]{\textcolor{red}{Gerardo: #1}}
\renewcommand{\jComment}[1]{\textcolor{green}{Jim: #1}}
\renewcommand{\rComment}[1]{\textcolor{magenta}{Ray: #1}}
\renewcommand{\lComment}[1]{\textcolor{purple}{Rolando: #1}}

\begin{document}

\title{Discrete Sliding
Symmetries, Dualities, and Self-dualities of 
Quantum Orbital Compass Models and $p+ip$ Superconducting Arrays}
\author{Zohar Nussinov}
\address{Theoretical Division,
Los Alamos National Laboratory, Los Alamos, NM 87545}
\author{Eduardo Fradkin}
\address{Department of Physics,
University of Illinois
1110 W. Green St., Urbana IL 61801-3080}

\date{Received \today }

\begin{abstract}
We study the spin-$1/2$ two and three dimensional Orbital Compass
Models relevant to the problem of orbital ordering in transition metal 
oxides. We show that these systems display 
self-dualities and novel (gauge-like) discrete sliding symmetries. 
An important and surprising consequence is that these models are dual
to (seemingly unrelated) recently studied models of $p+ip$ superconducting arrays. 
The duality transformations are constructed by means of a path-integral representation in 
discretized imaginary time 
and considering its ${\mathbb Z}_{2}$ spatial reflection symmetries
and space-time discrete rotations, we
obtain, in a transparent unified geometrical way, 
several dualities. We also introduce an alternative construction of the duality transformations using operator identities.
We discuss the consequences of these 
dualities for the order parameters and phase transitions of the orbital compass model and 
its generalizations, and apply these ideas 
to a number of related systems.
\end{abstract}


\maketitle

\section{Introduction} 

Orbital compass models offer a simple and qualitative description of the ordering of orbital degrees of freedom in a number of complex oxides such 
as the titanates \cite{ruthenates}. The degrees of freedom of these models describe 
the spatial orientation of the orbital degrees of freedom. Jahn-Teller
effects lead to anisotropic orbital compass like
interactions amongst the orbitals. When combined with 
the spin degrees of freedom, to which the orbitals are coupled via 
super-exchange in these systems \cite{KK} as well as by 
spin-orbit interactions, they lead to complex phase diagrams with phases that involve both spin and orbital 
 ordering (and disorder) to various degrees. Indeed, these systems offer an interesting 
 laboratory for the study of interesting anisotropic quantum nematic phases, with and without spin order, 
 and are a simple example of electronic liquid crystal phases \cite{ELC,jan}.

Orbital compass models also exhibit unusual and so far not well studied 
symmetries which play a big role in their physical properties. In the current article, 
we elucidate the discrete ``sliding" gauge-like
symmetries present in the two and three
dimensional orbital compass models. In two dimensions, these symmetries involve 
flipping the orbital degrees of freedom simultaneously along a single row or 
column of the lattice. These discrete symmetry transformations stand in-between the
 global symmetries familiar from spin systems and the local symmetries of gauge theories.  
 Although these are not truly gauge symmetries in the sense that they affect the boundary conditions,
 they are softer than the familiar global symmetries. In fact, for reasons discussed elsewhere \cite{bn} 
 these discrete sliding symmetries are alike gauge symmetries in that they cannot be spontaneously broken. 
 A direct consequence of the existence of these discrete sliding symmetries is that their natural order 
 parameters are {\em nematic}, 
which are invariant under
discrete sliding symmetries. Here we give an explicit construction of the nematic order parameters 
and potential physical consequences
are discussed. 

Gauge-like symmetries 
appear in a number of condensed matter systems. Exact local gauge symmetries are pervasive in the quantum hydrodynamics of incompressible and compressible quantum Hall systems, as a direct expression of their quantum hydrodynamics \cite{qhe}. Similarly, local gauge symmetries appear naturally in the context of strongly correlated systems such as the $t-J$ model, quantum dimer models, and other systems \cite{book}. 
Of particular interest for the problems discussed in this paper are the {\em sliding phases} 
of arrays of  Luttinger liquids \cite{SLL}, quantum Hall smectic (stripe) phases \cite{QHstripe}, 
DNA intercalates in lipid bilayers \cite{DNA}, 
as well as in some ring exchange models of frustrated antiferromagnets \cite{ring}.  
The discrete sliding symmetries we discuss here are a discrete, ${\mathbb Z}_2$, version of the 
{\em continuous} sliding symmetries of the systems mentioned above. The existence of 
sliding symmetries has profound effects on their quantum phase transitions, whose behavior
only begun to be understood quite recently \cite{lawler} and still remains largely unexplored. 

Amongst others, discrete sliding symmetries are present 
in spin \cite{oleg, henley,bt}, orbital 
\cite{Harris,BCN1,BCN2,NBCv,Mishra},
and superconducting array systems
\cite{Xu03,Xu04}. We further demonstrate
that the planar orbital
compass model \cite{BCN1,Mishra} 
and the Xu-Moore model \cite{Xu03,Xu04} of
two dimensional $p+ip$ superconducting arrays
are, in fact, one and the same system, related by 
a simple duality transformation.
Viewed in that light, the discrete sliding
symmetries which the Hamiltonians
describing superconducting arrays 
display are natural.
By applying our new dualities,
we find self-dualities
for the three dimensional orbital compass model
and several other systems.
These dualities do not rely on
operator representations \cite{fradkin78}, \cite{kogut}
nor on standard combinatorial loop/bond counting 
arguments or summation formulas \cite{standard*duality}. 
Rather, the new dualities that we
report here appear as simple geometrical reflections
between various spin and spatial axis. 
The dualities investigated in this paper
map such trivial geometrical reflection self-dualities in one model onto 
far less trivial weak-strong 
coupling self-dualities in other systems. 
In a formal setting, our dualities correspond to 
different space-time cuts
of a single classical action.
Choosing a certain time and spin
quantization axis, we find one spatial system
whilst choosing the time axis to lie
along another direction in space-time
leads to a seemingly very different 
(yet dual) spatial model. Our new, purely geometric, 
dualities further extend and complement, 
from a rather general perspective, 
the dualities generally derived via techniques
such as those in, e.g., \cite{Savit}.

The plan to the paper is as follows. In Section \ref{qp},
we introduce the planar orbital compass model in both
its isotropic and anisotropic incarnations. We
identify the many gauge like and 
single reflection symmetries of this model
(the latter reflection symmetry will,
as we will later find out, play the role
of self-duality).
In the aftermath, we construct order parameters 
invariant under these symmetries. 

In Section \ref{pps}, 
we discuss another two dimensional
XY system that possesses one dimensional
gauge like symmetries. This system 
has been argued to embody the quintessential
physics of a square lattice array of $p+ip$
superconducting grains. As we will show, 
this model is identical to the planar orbital
compass model.

In Section \ref{tdo}, we discuss the
three dimensional orbital compass system. 
This model has been considered to
embody the prototypical features of
orbital system Hamiltonian 
and might be directly relevant
to the so-called ``$t_{2g}$'' systems
(such as the vanadates and manganates)
in particular. We identify 
gauge like symmetries in this
system. As in the planar
case, we find  
nematic orders
invariant the gauge 
like symmetries. This in
turn suggests that orbital systems
might possess observable nematic orders.

In Section \ref{mc}, we 
employ simple geometrical reflections
to derive dualities for extended 
systems (now residing in
three dimensions). The dualities
of Xu and Moore  \cite{Xu03,Xu04} (derived
by Kramers-Wannier loop 
counting) form a subset of
the derived dualities.
The central actor in 
our scheme is a geometrical
inversion operator which 
allows us to set the 
imaginary time axis 
along different external 
space-time directions with similar
ideas for choosing the internal
spin quantization axis. These operations 
generate, in turn, many different 
dual models. With this geometrical
understanding of the observed
duality in hand, we return to
the self-dual point of 
\cite{Xu03,Xu04} and make comparisons
to other systems. 

In Section \ref{sec:operator}, we use an operator representation of the duality transformation
to rederive 
the dualities for the orbital
compass model which we obtained
via geometrical
reflections in the previous sections. 
Section \ref{conc} is devoted to the conclusions.
In Appendix \ref{cube} 
we discuss the self-duality of ``around the cube" models in transverse fields.

\section{Quantum Planar orbital-compass models: Symmetries.}
\label{qp}

We start with the planar compass-model. The compass models 
often serve as the simplest
caricatures for the physics of 3d orbital systems wherein 
Jahn-Teller interactions as well as magnetic exchange 
processes are dictated by the orientation of the orbitals
at the various lattice sites. 
In the orbital compass models,
the spin variables code for the orbital states. As orbitals extend
in real space, all orbital dependent interactions
are highly anisotropic- these interactions link the external
lattice directions with the internal ``spin''
(i.e. orbital) orientations. 
We refer the interested reader to 
\cite{orbital*models} where
the physics of orbital systems
and the orbital only models 
that we investigate is explored 
in depth.

The planar compass model is defined on the square lattice where at
each site~$\br$ there is a S=1/2 operator denoted 
by~$\bS_{\br} = \frac{\hbar}{2} \bf{\sigma}_{\br}$. 
The isotropic planar orbital model Hamiltonian 
\begin{equation}
\label{Hcompass}
H_{iso}=-J \sum_{\br} (\sigma^{x}_{\br} \sigma^{x}_{\br+\hat{e}_{x}} 
+ \sigma^{z}_{\br} 
\sigma^{z}_{\br+e_{z}}),
\end{equation}
where the nature of the interaction allows us to set $J>0$ \cite{AF}.

Unlike the more conventional nearest neighbor spin Hamiltonians
which posses a continuous global rotational symmetry,
the compass model Hamiltonian is not invariant under
arbitrary global rotations of all spins. [Physically, the lack 
of this symmetry is the direct consequence of 
the coupling between the internal polarization
directions (orbital states) and the external 
lattice directions (as much unlike spins, the orbitals
extend in real space) \cite{orbital*models}.] Instead, 
this model possesses many new non-trivial symmetries
corresponding to specific quantized angles of rotation of
all spins on given rows/columns and a single additional 
rather trivial reflection symmetry (which upon mapping 
will enable us to find a non-trivial
weak to strong coupling self-duality
in another model). As a consequence of these symmetries, 
this model harbors an infinite degeneracy of all states 
and of its ground states in particular. Let us consider the system with 
open boundary conditions on an $L \times L$ 
square lattice and let us define an operator on an arbitrary horizontal line 
(of ordinate z)
$\hat{O}_{z} = \prod_{x=-L}^{L} \sigma^{z}_{x}$ 
and an operator on an arbitrary vertical line 
(of horizontal intercept x)  
$\hat{O}_{x} = \prod_{z=-L}^{L} \sigma^{x}_{z}$. It is readily verified that
for all sites $\vec{r}$ whose z component is $r_{z} = z $, the product 
$\hat{O}_{z}^{-1} \sigma^{x}_{\vec{r}} \hat{O}_{z} = - \sigma^{x}_{\vec{r}}$
while $\hat{O}_{z}^{-1} \sigma^{z}_{\vec{r}} \hat{O}_{z} =  
\sigma^{z}_{\vec{r}}$.
Similarly, $\hat{O}_{x}$ inverts the z component of all spins on 
a vertical line,
while leaving $\sigma_{x}$ untouched. In the case of symmetric
exchange constants for bonds along the x and z axis,  
as in the compass model under consideration here, 
(where both exchange constants are equal
to $J$), we further have a single additional 
${\mathbb Z}_2$ reflection symmetry ($\sigma_{x} \to \sigma_{z}$, 
$\sigma_{z} \to 
\sigma_{x}$)- a rotation by $\pi$ about the symmetric line (the 45 degree line
in the xz plane), i.e. 
$\hat{O}_{Reflection} = \prod_{\vec{r}} \exp[i \frac{\pi \sqrt{2}}{4} 
(\sigma^{x}_{\vec{r}} + \sigma^{z}_{\vec{r}})]$. For each of these
operations, $\hat{O}_{\alpha}^{-1} H \hat{O}_{\alpha} = H$. This 
symmetry, $\hat{O}_{Reflection}$ is a manifestation of self-duality
present in the model- we will explain the origin of this comment later.
Putting all of the pieces together, 
as a consequence of these
symmetries, each state is, at least, ${\cal{O}}(2^{L})$ degenerate. 
Formally, these symmetries constitute a gauge like symmetry which is 
intermediate between a local gauge symmetry (whose volume scales as
the system area) and a global gauge symmetry  (whose logarithmic volume
is point-like). These intermediate gauge symmetries suggest that 
non-trivialities may occur. As it turns out, such large discrete symmetries
do not prohibit ordering in classical variants of this model albeit 
complicating matters significantly \cite{BCN1,NBCv,Mishra}. 
This ordering tendency may be expected
to become fortified by quantum fluctuations 
(``quantum order out of disorder'') \cite{moessner}. 

As an aside, we note that the global nematic 
symmetries (the global rotation of all spins by 
$n \pi/2$ with $n=0,1,2,3$) are
not independent symmetries on top
of the gauge like symmetries 
discussed above. Rather, there
is only one ($\sigma_{x} \to  \sigma_{z}$, $\sigma_{z} \to 
\sigma_{x}$)- a rotation by $\pi$ about the symmetric line 
(the 45 degree line in the xz plane)) additional symmetry 
supplanting the gauge like symmetries. 
To see this, first note that the global inversion operation,
$\vec{\sigma} \to - \vec{\sigma}$, is a composite of
the row/column inversion symmetries: An inversion of 
$\sigma_{x}$ on all rows followed by an inversion
of $\sigma_{z}$ on all columns leads to 
the global inversion operation. Next, note that by fusing
the global inversion symmetry with the global ${\mathbb Z}_2$
reflection symmetry, we may produce the four global 
nematic symmetry operations (rotations by $n \pi/2$).
Thus, unlike what is suggested by \cite{Mishra},
the global nematic symmetries do not supplant the
gauge-like symmetries and no less 
important, the quantum system
possesses the above reported
gauge like symmetries embodied
by the operators $\hat{O}_{x,z}$.

The classical (large $S$) ground state sector of the orbital compass model
further possesses
an additional continuous ($U(1)$) symmetry not captured by the discrete 
$({\mathbb Z}_2)^{2L+1}$ symmetries ($2^{L}$  of these associated with 
horizontal, $2^{L}$ associated with vertical discrete 
spin flip symmetries, and one ${\mathbb Z}_2$ symmetry 
being the ($\sigma_{x} \to \sigma_{z}$, $\sigma_{z} \to 
\sigma_{x}$) reflection symmetry) detailed above. 
This continuous symmetry is
made evident by noting that any constant spin-field, 
${\bf \sigma_{\br}}={\bf \sigma}$, is a ground state.
First, we note that 
$\sum_{\alpha=x,z} [\sigma_{\br}^{(\alpha)}]^2$ is constant. 
Thus, up to an irrelevant constant, the general Hamiltonian of 
Eq.~\eqref{Hcompass}
is \begin{equation}
\label{Hdiff}
H_{iso}^{cl}=\tfrac J2\sum_{\br,\alpha}\bigl
(\sigma_{\br}^{(\alpha)}-\sigma_{\br+\hate_\alpha}^{(\alpha)}\bigr)^2,
\end{equation}
which is obviously minimized when the spin field is constant.
We emphasize that the continuous symmetries which underscore these ground 
states are just symmetries of the states and \emph{not} of the Hamiltonian 
itself. In common parlance, these are {\em emergent symmetries}
specific only to the ground state sector.
Therefore, at least in the orbital-only models, we are not in a setting 
where a Mermin-Wagner argument can be applied.

With an eye towards things to come, let us 
now introduce and examine the anisotropic
planar compass model,
\begin{eqnarray}
H = - J_{x} \sum_{\bf r} \sigma^{x}_{\bf r} \sigma^{x}_{\bf r + \hat{e}_{x}}
-   J_{z} \sum_{\bf r} \sigma^{z}_{\bf r} \sigma^{z}_{\bf r + \hat{e}_{z}}.
\label{an2d}
\end{eqnarray}

It is readily verified that this more general Hamiltonian
harbors all of the one dimensional gauge like symmetries 
encapsulated by $\hat{O}_{x,z}$. The only symmetry which 
does not persist for arbitrary $J_{x}, J_{z}$ is the reflection symmetry. 
[Insofar as its underlying physics is concerned, this anisotropic compass 
model emulates Jahn-Teller distortions on a strained lattice 
\cite{orbital*models}.] 

The two terms in the anisotropic compass model of Eq.(\ref{an2d}),
trivially compete. The first term favors ordering of the spins
parallel to the x axis while the second 
favors
an ordering of the spins parallel to the z-axis.
Order becomes more inhibited when the competition between the
two terms becomes the strongest ($J_{x} =  \pm J_{z}$)
as it indeed occurs within the compass
model of Eq.(\ref{Hcompass}). We note that the gauge like symmetries 
(encapsulated by the column/row $\hat{O}_{x,z}$ generators) 
preserve the Hamiltonian also for arbitrary $|J_{x}| \neq |J_{z}|$.
A natural (smectic-like) order parameter in the  orbital compass model  
monitors the tendency of the spins to order along their preferred directions
\begin{eqnarray}
m = \langle \sigma^{x}_{\br} \sigma^{x}_{\br+\hat{e}_{x}} - \sigma^{z}_{\br} 
\sigma^{z}_{\br+\hat{e}_{z}} \rangle.
\end{eqnarray}
(Just as in smectic liquid crystals, having all spins point in the 
$\hat{e}_{\alpha}$ direction
or in the ($- \hat{e}_{\alpha}$) direction is one and the same
insofar as the above order parameter is concerned). 
This nematic like order parameter is
invariant under all gauge like
symmetries.

Similar to the xy symmetric order parameter
above, for the anisotropic planar orbital compass model (say $|J_{x}| > 
|J_{z}|$),
we may consider the Ising like nematic order parameter,
\begin{eqnarray}
m_{x} = \langle \sigma^{x}_{\br} \sigma^{x}_{\br+\hat{e}_{x}}  \rangle,
\end{eqnarray}
with a similar definition for the system with $|J_{z}| > |J_{x}|$. 
These order parameters are invariant under the 
gauge-like symmetries of the system.

The classical, large $S$, rendition of this 
model, has similar nematic like
order parameters invariant under all
gauge-like symmetries \cite{BCN1, NBCv,Mishra}. 
Here, and in fact for all spins $S > 1/2$,
the order parameter can be 
local (not a bond order
parameter involving 
two spins). All quantities
$Q_{\alpha \beta} = S^{\alpha} S^{\beta} - \frac{1}{d} \delta_{\alpha \beta}$,
with $\alpha,\beta$ internal spin indices, and
with $d$ the dimension ($d=2$ in the planar
higher spin extensions of the orbital compass model) 
are invariant under all gauge like
symmetries. The order parameter $\langle Q_{11} \rangle$
is anticipated for $|J_{x}| > |J_{z}|$ (and $\langle Q_{22} \rangle$
for $|J_{z}| > |J_{x}|$). Similar quantities may be 
introduced for higher dimensional ($d>2$) generalizations 
of the planar compass model.

\section{$p+ip$ Superconducting Arrays}
\label{pps}

A Hamiltonian describing a square lattice of $p+ip$
superconducting grains (e.g. Sr$_{2}$RuO$_{4}$) 
was recently suggested \cite{Xu03,Xu04}, 
\begin{eqnarray}
H = - K \sum_{\Box} \sigma^{z} \sigma^{z} \sigma^{z} \sigma^{z}
- h \sum_{\bf r} \sigma_{\bf r}^{x}.
\label{XM}
\end{eqnarray}
Here, the four spin product is the product of
all spins residing at the four vertices of a given plaquette $\Box$ (not on its bonds as for gauge fields!).
As noted by Xu and Moore, \cite{Xu03}, the quantity
\begin{eqnarray}
\hat{O}_{P} = \prod_{\bf r} \sigma^{x}_{\bf r},
\end{eqnarray}
with the string product (along ``P'') extending over all spins in a
given row ($r_{z} = z$) or a given column ($r_{x} =x$),
is conserved. The discrete (gauge-like) sliding symmetry of 
this model is similar to that of
the planar orbital compass
model and we will indeed
show that these two models 
are actually dual to each other.

The central derivation in 
\cite{Xu03,Xu04} was a self-duality
of the Hamiltonian in Eq.(\ref{XM}) via a tour de
force Wannier Kramers loop
counting arguments. The form
of this self-duality is somewhat similar
(yet still very different) to
the beautiful self-dualities
of \cite{amit}. Similar dualties were discussed
in the ring exchange systems of \cite{ring}. 
In these models not only a relation amongst strong
and weak coupling is given
by the self-duality but the self-duality further
intertwines the various 
terms (e.g. large $h$ is related
to small $K$ in the self-duality 
of Eq.(\ref{XM}) and vice versa 
as found by Xu and Moore).

We will shortly 
establish that the rather complicated looking
weak coupling to strong coupling self-duality 
of Eq.(\ref{XM}) derived by \cite{Xu03,Xu04} 
immediately follows from a 
very simple purely geometric
(${\mathbb Z}_{2}$ reflection) self-duality
of the planar orbital compass
model. This self-duality may
also be related (albeit in a 
less general fashion) to the trivial
geometrical self-duality
of the planar orbital compass
model via the operator representations 
of Section(\ref{sec:operator}).
In the aftermath, the plaquette
coefficient $K$ in Eq.(\ref{XM})
may be related to the exchange
amplitude $J_{x}$ of Eq.(\ref{an2d})
whereas the transverse magnetic field
$h$ of Eq.(\ref{XM}) becomes trivially
related to $J_{z}$ of Eq.(\ref{an2d}).

\section{Symmetries of the three dimensional Orbital Compass Model}
\label{tdo}

The canonical prototype of all orbital-spin \cite{KK} and orbital-orbital 
interactions is the orbital compass model \cite{brink03}. The model 
is defined on the cubic 
lattice where at each site~$\vec{r}$ there is an S=1/2 operator 
denoted by~$\vec{S}_{\vec{r}} = \frac{\hbar}{2} \vec{\sigma}_{\vec{r}}$. 
The orbital model Hamiltonian 
\begin{equation}
\label{Hcompass*}
H=J 
\sum_{\vec{r}} (\sigma^{x}_{\vec{r}} \sigma^{x}_{\vec{r}+\hat{e}_{x}} 
+ \sigma^{y}_{\vec{r}} \sigma^{y}_{\vec{r}+\hat{e}_{y}}
+ \sigma^{z}_{\vec{r}} 
\sigma^{z}_{\vec{r}+e_{z}}).
\end{equation}

Let us define an operator on an arbitrary xy plane P (of intercept z)
$\hat{O}_{P;z} = \prod_{\vec{r} \in P} \sigma^{z}_{\vec{r}}$ 
with similar definitions for $\hat{O}_{P;x}$ and 
$\hat{O}_{P;y}$. These operators may be recast
as rotations by $\pi$ about an axis orthogonal 
to the plane. For all sites $\vec{r}$ in the xy plane $P$ 
whose z component is $r_{z} = z $, up to a multiplicative phase factor,
the operator $\hat{O}_{P;z} =  \exp[i (\pi/2) \sigma^{z}_{P}/\hbar]$
with $\sigma^{z}_{P} = \sum_{\vec{r} \in P} \sigma^{z}_{\vec{r}}$.
The products $\hat{O}_{P;z}^{-1} \sigma^{P;x,y}_{\vec{r}} 
\hat{O}_{P;z} = - \sigma^{x,y}_{\vec{r}}$ while
$\hat{O}_{P;z}^{-1} \sigma^{z}_{\vec{r}} \hat{O}_{P;z} =  
\sigma^{z}_{\vec{r}}$.
Similarly, $\hat{O}_{P;x}$ inverts the y and z 
component of all spins on the yz plane 
of intercept $x$ while leaving $\sigma_{x}$ untouched. 
These ``string'' operators spanning the entire plane
commute with the Hamiltonian,
$[H, \hat{O}_{P;\alpha}]=0$. The classical orbital compass model 
has an exact $[{\mathbb Z}_2]^{3L^{2}}$ symmetry 
(along each chain parallel to the cubic $\alpha$ ($x,y,$ or z)
axis, we may reflect the $\alpha$ spin component, 
$S_{\alpha} \to -S_{\alpha}$, while keeping all 
other spin components unchanged, 
$S_{\beta \neq \alpha} \to S_{\beta \neq \alpha}$).
The quantum orbital compass model has 
a lower exact $[{\mathbb Z}_2]^{3L}$ gauge like 
symmetry (forming a subset of the
larger $[{\mathbb Z}_2]^{3L^{2}}$ symmetry
present for classical spins).
As alluded to above, the gauge like $[{\mathbb Z}_2]^{3L}$ symmetries
of this quantum $S = 1/2$ case
(as well as all representations),  
become evident once we rotate, with no change ensuing 
in the Hamiltonian, all spins in a plane orthogonal to
the cubic lattice direction $\alpha$ by $\pi$ about the 
internal $S_{\alpha}$ quantization axis. 

As before, let us now introduce and examine 
the anisotropic orbital compass model,
\begin{eqnarray}
H =&& \!\!\!\!\! - J_{x} \sum_{\bf r} \sigma^{x}_{\bf r} \sigma^{x}_{\bf r + \hat{e}_{x}}
 - J_{y} \sum_{\bf r} \sigma^{y}_{\bf r} \sigma^{y}_{\bf r + \hat{e}_{y}}
 - J_{z} \sum_{\bf r} \sigma^{z}_{\bf r} \sigma^{z}_{\bf r + \hat{e}_{z}}.
 \nonumber \\
 &&
\label{an3d}
\end{eqnarray}
(The isotropic orbital compass model corresponds
to $J_{x,y,z} = -J$). The anisotropic orbital compass
model possesses all of the gauge like symmetries
of the isotropic orbital compass model (planar
rotations in the quantum model and more numerous single 
line inversions in the classical case). Further, if
at least any two of the three exchange constants $\{ J_{\alpha} \}$ 
are identical the system possess a reflection symmetry.

As in the planar orbital compass model, nematic like 
order parameters may be constructed for both the 
isotropic and anisotropic systems. Thus, we
naturally predict the existence of observable 
nematic orbital orders in $t_{2g}$ systems.

In what follows, we will also investigate
a related system governed by the 
Hamiltonian
\begin{eqnarray}
H =&& \!\!\!\!\! - J_{x} \sum_{\bf r} \sigma^{x}_{\bf r} 
\sigma^{x}_{\bf r + \hat{e}_{x}}
 - J_{y} \sum_{\bf r} \sigma^{z}_{\bf r} \sigma^{z}_{\bf r + \hat{e}_{y}}
 - J_{z} \sum_{\bf r} \sigma^{z}_{\bf r} \sigma^{z}_{\bf r + \hat{e}_{z}}.
 \nonumber \\
 &&
\label{an3d1}
\end{eqnarray}
Note the similarity between the XY model of Eq.(\ref{an3d1}) and the
orbital compass model of Eq.(\ref{an3d}). In the limit
$J_{z} =0$, Eq.(\ref{an3d1}) trivially degenerates
into the strained planar orbital compass
model of Eq.(\ref{an2d}).  

We will construct new ``plaquette models''
(in which the spins reside on the lattice
sites not on bonds) dual
to Eq.(\ref{an3d1})
which possess a self-duality and gauge like
symmetry, naturally extending the 
results of \cite{Xu03,Xu04}.

\section{New dualities and self-dualities found by 
planar reflections}
\label{mc}

We now transform the zero temperature Quantum problem of 
Eq.(\ref{Hcompass}) onto a classical problem 
in $(d+1)$ dimensions. To this end, we
work in a basis quantized along 
$\sigma^{z} (= \pm 1) $. 
We now consider the basis spanned
by two spins $(\sigma^{z}, \sigma^{z \prime})$
at the same spatial site ${\bf r}$ yet at two consecutive imaginary 
times $\tau$ and $(\tau+ \Delta \tau)$.
The transfer matrices corresponding to $\alpha e^{\overline{h} \sigma^{x}}$
(stemming, in the imaginary time
formalism from a propagator~ $e^{-H \Delta \tau}$ 
such as  $e^{h \sigma^{x} \Delta \tau}$ with $\overline{h} \equiv 
h \Delta \tau$) 
and $e^{\overline{J} \sigma^{z} \sigma^{z \prime}}$
(or, with space time coordinates explicitly instated, 
$e^{\overline{J} \sigma^{z}_{\bf r, \tau} 
\sigma^{z}_{\bf r, \tau + \Delta \tau}}$) are the same 
provided that $\tanh  \overline{h} = e^{-2 \overline{J}}$ 
(or equivalently $\sinh 2 \overline{h} \sinh 2 \overline{J} =1)$
and $\alpha = (2 \sinh \overline{J})^{1/2}$. Similarly,
the non-vanishing eigenvalues of
the transfer matrices 
\begin{eqnarray} 
\exp[ K_{x} \sigma^{z}_{i,\tau}
\sigma^{z}_{i+1,\tau} \sigma^{z}_{i,\tau+ \Delta \tau} 
\sigma^{z}_{i+1,\tau}] 
\label{Kxt}
\end{eqnarray}
and
\begin{eqnarray}
\exp[ \overline{J}_{x} \sigma^{x}_{i,\tau} \sigma^{x}_{i+1, \tau}]
\label{Jxt}
\end{eqnarray}
are equivalent once 
$\sinh 2 K_{x} \sinh 2 \overline{J}_{x} =1$.

In the standard imaginary time 
mapping of quantum systems to 
classical actions, we identify $\overline{J}_{\alpha} = J_{\alpha} 
\Delta \tau$
with the aforementioned 
$\Delta \tau$ the lattice spacing along the imaginary time direction.

The generalized classical Euclidean action corresponding 
to Eq.(\ref{Hcompass}) is 
\begin{eqnarray}
S =&&\!\!\!\! - K_{x} \sum_{\Box \in (x \tau) ~ \mbox{plane}} \sigma^{z}_{\bf r,\tau} 
\sigma^{z}_{\bf r, \tau + \Delta \tau} \sigma^{z}_{\bf r + \hat{e}_{x}, \tau}
\sigma^{z}_{\bf r + \hat{e}_{x}, \tau + \Delta \tau} 
\nonumber \\
&& -
(\Delta \tau) J_{z} \sum_{\bf{r}} \sigma^{z}_{\br} 
\sigma^{z}_{\br+e_{z}}
\nonumber \\
&&
\end{eqnarray}

\begin{figure}
\centerline{\psfig{figure=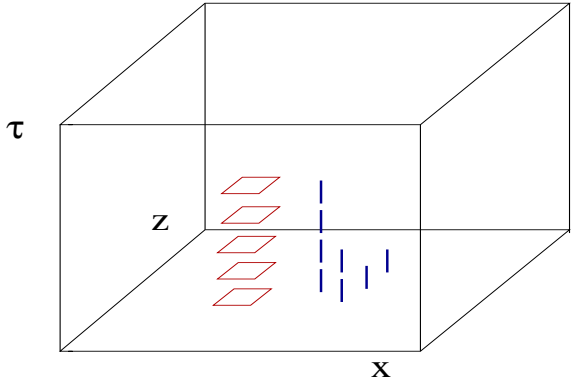,height=5.0cm,width=7.5cm,angle=0}}
\caption{The classical Euclidean action corresponding
to the Hamiltonian of Eq.(\ref{XM}) at zero temperature
in a basis quantized along the $\sigma^{z}$ direction. The transverse
field leads to bonds parallel
to the imaginary time axis while
the four plaquette interactions become
replicated along the imaginary time axis.
Taking an equal time slice of this system
we find the four spin term of Eq.(\ref{XM})
and the on-site magnetic field
term. If we interchange $\tau$ with $z$,
we find the anisotropic planar orbital compass model of Eq.(\ref{an2d})
in the basis quantized along the $\sigma^{x}$ direction.}
\label{FIG:DUAL1}
\end{figure}

A schematic of this action in Euclidean space-time
is shown in Fig.(\ref{FIG:DUAL1}).
If we relabel the axes and replace the spatial index $x$ with the
temporal index $\tau$, we will immediately find the classical action 
corresponding to the the Hamiltonian of Eq.(\ref{XM}) depicting
$p+ip$ superconducting grains in a square grid. This trivially
suggests that the anisotropic planar orbital compass system 
(Eq.(\ref{an2d})) and the Xu-Moore Hamiltonian (Eq.(\ref{XM})) 
are dual to each other. In Section(\ref{sec:operator}), we sketch a detailed 
derivation of this duality by the operator dualities of 
\cite{fradkin78}, \cite{kogut}. This classical action
follows from the equivalence of the transfer matrices
corresponding to Eqs.(\ref{Kxt},\ref{Jxt}) or, alternatively,
from the equivalence of Eq.(\ref{an2d}))  to 
Eq.(\ref{XM}) [which will be proved in detail by operator
representations in Section(\ref{sec:operator})]
and the relation between the transfer
matrices corresponding to $e^{\overline{h} \sigma^{x}}$
and $e^{\overline{J} \sigma^{z} \sigma^{z \prime}}$.

We find that the classical 
action corresponding to the 
model of Eq.(\ref{an3d1}) is
\begin{eqnarray}
S = [- \tanh^{-1}(e^{-2 J_{x} \Delta \tau}) \sum_{\Box \in x 
 \tau ~\mbox{plane}} \sigma \sigma \sigma \sigma 
\nonumber
\\ - \Delta \tau J_{z} \sum_{z~ \mbox{direction}} \sigma \sigma \nonumber
\\ - \tanh^{-1}(e^{-2J_{y} \Delta \tau}) \sum_{\Box \in y 
\tau ~\mbox{plane}} \sigma \sigma \sigma \sigma].
\label{S3d-direct}
\end{eqnarray}
Here and elsewhere, $\sigma = \pm 1$ are c-numbers 
and we omit the (z) polarization superscripts. 

We now extend the duality of self-duality
of Eq.(\ref{XM}) to the three dimensional
arena. First note that by 
interchanging the imaginary time
coordinate $\tau$ with the spatial z coordinate, 
we find that
\begin{eqnarray}
H = -(K_{xz} \sum_{\Box \in xz} \sigma^{z} \sigma^{z} \sigma^{z} \sigma^{z} 
\nonumber
\\ + K_{yz} \sum_{\Box \in yz} \sigma^{z} \sigma^{z} \sigma^{z} \sigma^{z}  
+ h \sum_{\bf r} \sigma^{x}_{\bf r})
\label{XMnew}
\end{eqnarray}
is dual to the 
system given by the Hamiltonian of Eq.(\ref{an3d1}). 
Let us now derive self-dualities of this extended
three dimensional system (en passant, effortlessly proving
the central result of \cite{Xu03,Xu04}).
  
Expressing the action corresponding to
the Hamiltonian of Eq.(\ref{an3d1}) in a spin 
eigen-basis of $\sigma^{x}$ and inverting the 
spatial $z$ and $x$ coordinates of any
site ${\bf r}$ (a ${\mathbb Z}_2$ operation), we obtain 
\begin{eqnarray}
S_{dual} = [ -\tanh^{-1} (e^{-2J_{z} \Delta \tau}) \sum_{\Box
\in x 
\tau ~\mbox{plane}} \sigma \sigma \sigma \sigma 
\nonumber
\\ - \Delta \tau J_{x} \sum_{z~ \mbox{direction}} \sigma \sigma \nonumber
\\ - \tanh^{-1}(e^{-2J_{y} \Delta \tau}) \sum_{\Box \in y 
\tau ~\mbox{plane}} \sigma \sigma \sigma \sigma].
\label{S3d-dual}
\end{eqnarray}

Looking at Eqs.(\ref{S3d-direct},\ref{S3d-dual}), 
we see that if 
\begin{eqnarray}
S = [A \sum_{\Box \in x 
\tau ~\mbox{plane}} \sigma \sigma \sigma \sigma 
\nonumber
\\ + B \sum_{z~ \mbox{direction}} \sigma \sigma \nonumber
\\ + C \sum_{\Box \in y 
\tau ~\mbox{plane}} \sigma \sigma \sigma \sigma]
\label{S3d-direct*}
\end{eqnarray}
then 
\begin{eqnarray}
S_{dual} = [\tilde{A} \sum_{\Box \in x 
\tau ~\mbox{plane}} \sigma \sigma \sigma \sigma 
\nonumber
\\ + \tilde{B} \sum_{z~ \mbox{direction}} \sigma \sigma \nonumber
\\ + \tilde{C} \sum_{\Box \in y 
\tau ~\mbox{plane}} \sigma \sigma \sigma \sigma].
\label{S3d-dual*}
\end{eqnarray}

Here, $A = - \tanh^{-1}(e^{-2 \overline{J}_{x}})$, $B = - \overline{J}_{z}$,
and $C = - \tanh^{-1}(e^{- 2 \overline{J}_{y}})$. Similarly, 
$\tilde{A} = - \tanh^{-1}(e^{-2\overline{J}_{z}})$, $\tilde{B} = 
- \overline{J}_{x}$, and $\tilde{C} = C$. 
Eliminating $\overline{J}_{x,z}$,
we find that 

\begin{eqnarray}
\sinh 2\tilde{A} \sinh 2B = 1, \nonumber
\\ \sinh 2A \sinh 2 \tilde{B} = 1, \nonumber
\\ \tilde{C} = C.
\end{eqnarray}

Taken together, these relations
imply that 
\begin{eqnarray}
\sinh 2 A \sinh 2B = 1
\end{eqnarray}
is a self-dual line
for any value of $C$. 

This extends the dualities
of \cite{Xu03,Xu04} in a very natural fashion
to higher dimensions. Moreover,
note in this formalism the dualities
just ``fall into our lap''-
no involved calculations nor loop counting
were necessary. The duality is
a trivial geometrical reflection.

\section{
Operator duality transformations of the Quantum Planar orbital-compass models onto quantum Ising plaquette models}
\label{sec:operator}

A duality between the planar orbital compass model
(Eq.(\ref{an2d}))  
and the Xu-Moore model of Eq.(\ref{XM}) 
is suggested by the cubic point group 
operation interchanging $x$ with $\tau$
in Fig.(\ref{FIG:DUAL1}). We now prove this duality
at all temperatures. To this end, we invoke a simple operator duality
transformation followed by a summation over the horizontal bonds
(which amounts to a standard gauge fix) in the model that results. 
The upshot of the up and coming discussion is 
that the quantum planar compass model of Eqs.(\ref{Hcompass})
can be mapped onto the Hamiltonian of Eq.(\ref{XM}) precisely
at its zero temperature self-dual point.

\begin{figure}
\centerline{\psfig{figure=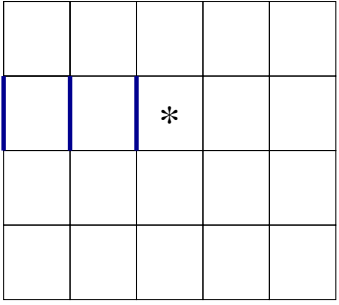,height=4.0cm,width=4.5cm,angle=0}}
\caption{A graphical representation 
of $\sigma^{x}$ the string product of $\tau^{z}$ on all vertical 
bonds from the  boundary up to the dual lattice site.} \label{FIG:SETUP2}
\end{figure}

\begin{figure}
\centerline{\psfig{figure=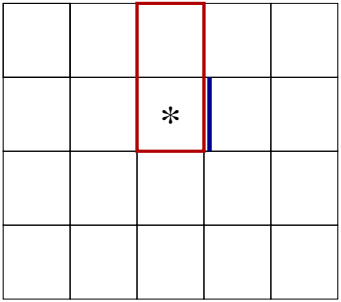,height=4.0cm,width=4.5cm,angle=0}}
\caption{The product of $J_{x} \sigma^{x}_{\bf r} \sigma^{x}_{\bf r 
+ \hat{e}_{x}}$
becomes $J_{x} \tau^{z}$ on a single vertical bond on the right-hand side of
the dual plaquette site corresponding to ${\bf r}$.
The product $J_{z} \sigma^{z}_{\bf r} \sigma^{z}_{\bf r + \hat{e}_{z}}$
becomes in the dual representation the product of
all $\tau^{x}$ operators forming the outer shell of a 
vertical domino multiplied by $J_{z}$.
Putting all of the pieces together, the Hamiltonian 
becomes the sum of $J_{z}$ multiplying a domino shell
of $\tau^{x}$ on bonds augmented by $J_{x}$ multiplying a
single vertical bond on which $\tau^{z}$ is placed.}. 
\label{FIG:SETUP5}
\end{figure}

\begin{figure}
\centerline{\psfig{figure=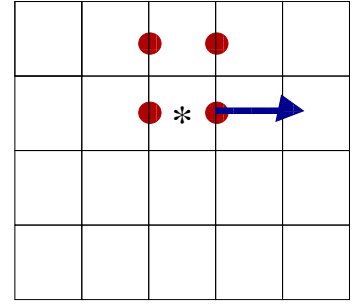,height=4.0cm,width=4.5cm,angle=0}}
\caption{Choosing a gauge in which $\tau^{x} =1$ on all
horizontal bonds, identifying the centers
of the vertical bonds as sites, we find that the Hamiltonian 
corresponds to the product of four $\sigma^{x}$ operators
on the vertices of a plaquette 
$[Ks^{x}_{\bf{r^{*}}} s^{x}_{\bf{r^{*}+ \hat{e}_{z}}}
s^{x}_{\bf r^{*} + \hat{e}_{z}- \hat{e}_{x}} s^{x}_{\bf r^{*} - \hat{e}_{x}}]$ 
augmented by an 
external transverse field giving rise to $h s^{z}_{\bf r ^{*}}$.
Here, $h= J_{x}$ and $K = J_{y}$. Thus, 
the planar orbital compass model is
dual to the superconducting array system
of \cite{Xu03,Xu04}.}. 
\label{FIG:SETUP6}
\end{figure}

The salient feature of the Pauli matrices 
$\sigma^{x}$
and $\sigma^{z}$ is that they
anti-commute at a common site 
while commuting everywhere 
else. 
It is readily verified
\cite{fradkin78},\cite{kogut}
that these relations are
preserved by the canonical
duality relations on the dual lattice
\begin{eqnarray}
\sigma^{z}_{\bf r}  = \tau^{x} \tau^{x} \tau^{x} \tau^{x}
\label{stx}
\end{eqnarray}
with the plaquette product of $\tau^{x}$ on the right hand side 
corresponding to the four spins surrounding the dual 
lattice site ${\bf r^{*}}$ (the center of the 
plaquette as shown in Fig.(\ref{FIG:SETUP1}) below),
and 
\begin{eqnarray}
\sigma^{x}_{\bf r} = \prod_{x \le x^{*}} \tau^{z}_{x,x+\hat{e}_{z}},
\label{stz}
\end{eqnarray}
the product of $\tau^{z}$ placed along vertical bonds 
(linking x and $x+\hat{e}_{z}$) along
a horizontal line- see Fig.(\ref{FIG:SETUP2}). 
The series of transformations below
leading to Eq.(\ref{2plaq*})
may be vividly followed 
in Figures(\ref{FIG:SETUP5}-\ref{FIG:SETUP6}).

Inserting Eqs.(\ref{stx},\ref{stz}) into Eq.(\ref{Hcompass}), 
we obtain 
\begin{eqnarray}
H_{iso}=-J \sum_{\br^{*}} [\tau^{z}_{x^{*}+\hat{e}_{x},x^{*}+\hat{e}_{x}+\hat{e}_{z}} \nonumber
\\
+ \sum_{\br^{*}} \tau^{x}_{\bf{r^{*}, r^{*}+\hat{e}_{z}}} \tau^{x}_{\bf{r^{*}+2 \hat{e}_{z}, 
r^{*}+2\hat{e}_{z}}} 
\tau^{x}_{\bf r^{*}+ 2 \hat{e}_{z},r^{*}+ 2 \hat{e}_{z} - \hat{e}_{x}} \nonumber
\\
\tau^{x}_{\bf r^{*} + \hat{e}_{z}, r^{*} + \hat{e}_{z} - \hat{e}_{x}}
\tau^{x}_{\bf r^{*}  + \hat{e}_{z} - \hat{e}_{x}, r^{*} - \hat{e}_{x}}
\tau^{x}_{\bf r^{*} - \hat{e}_{x}, r^{*}}].
\label{2plaq}
\end{eqnarray} 
The first term corresponds to an external transverse 
magnetic field of strength $J$ along the z-axis acting
on all vertical bonds  
while the second term encapsulates the product
of six bonds forming the outer shell of two plaquettes pasted together
along the z-axis. The bond common to the two plaquettes evaporated due
to the relation $\tau_{x}^{2}=1$. The net result of 
Eq.(\ref{2plaq}) is shown in Fig.(\ref{FIG:SETUP5}).

Next, we choose the longitudinal gauge wherein
all horizontal bonds have $\tau^{z}=1$.
This can be achieved via explicit gauge transformations
or by simply noting that in the representation with horizontal 
bonds with $\tau^{z}=1$, the duality relations
of Eqs.(\ref{stx}, \ref{stz}) become identical 
to the duality relations in one dimensional 
spin chains (performed independently for each horizontal row) which trivially
satisfy the commutation of spin variables of
different sites, the anticommutation of the x and z
components of the spin on the same site and the
square of each spin operator. In this longitudinal gauge where 
$\tau^{x}_{{\bf r^{*}}{\bf r^{*}} + \hat{e}_{x}} =1$, 
the Hamiltonian now involves only vertically oriented bonds (parallel to
the z-axis). Defining spins $s^{\alpha}_{\br^{*}} = \tau^{\alpha}_{\bf r^{*}, 
r^{*}+ \hat{e}_{z}}$,
\begin{eqnarray}
H_{iso}= 
-K \sum_{\br^{*}} s^{x}_{\bf{r^{*}}} s^{x}_{\bf{r^{*}+ \hat{e}_{z}}}
s^{x}_{\bf r^{*} + \hat{e}_{z}- \hat{e}_{x}}
s^{x}_{\bf r^{*} - \hat{e}_{x}} \nonumber
\\
-h \sum_{\br^{*}} s^{z}_{\br^{*}},
\label{2plaq*}
\end{eqnarray} 
with the new parameters $h=K$
being equal to the former $J$
of Eq.(\ref{2plaq}). Thus
the isotropic planar
compass orbital model
lies precisely on 
the zero temperature self-dual 
line $h=k$ of Eq.(\ref{2plaq*}).
This result is shown in Fig.(\ref{FIG:SETUP6}).

\begin{figure}
\centerline{\psfig{figure=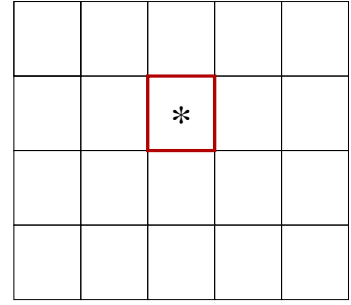,height=4.0cm,width=4.5cm,angle=0}}
\caption{The lattice and a dual lattice site (marked by an
asterisk ``*'' at a plaquette center). Here we illustrate the representation
of $\sigma^{z}$ as on a given dual plaquette
site as the product of $\tau^{x}$ operators
placed on all 4 bonds composing the plaquette
of the original lattice. } \label{FIG:SETUP1}
\end{figure}

\section{Conclusions}
\label{conc}

In conclusion, we investigated 
several systems displaying discrete (gauge-like) sliding symmetries and illustrated that
two such systems are dual to 
each other. The enhanced discrete sliding symmetries in these systems
go hand in hand with a dimensional reduction 
that occurs in several limiting
cases of these systems (e.g., 
the system of Eq.(\ref{XM}) 
in the limit of $h =0$
is none other than 
1+1 dimensional version
of the one dimensional 
Ising model.)
We found nematic order parameters
invariant under these symmetries. This suggests the
specter of detectable orbital nematic
orders in $t_{2g}$ orbital systems.

The superconducting array model
of \cite{Xu03}, \cite{Xu04}
is dual to the planar orbital 
compass model and as
such has a finite temperature
transition for a large $S$ 
incarnation at its self-dual point.

We find that dual models may be derived by flipping the spatial and 
imaginary time axis 
(and/or quantization axis). 
In an upcoming work,
we will elaborate on 
this novel approach to
dual models as different cuts
of a higher dimensional theory \cite{zn}.

The nature of the quantum phase transitions in these systems remains an open problem. 
A straightforward examination of the Xu-Moore model shows that the finite temperature 
transition from a high temperature disordered phase to a low temperature phase in which 
the product of Ising spins on pairs of sites belonging to different sub-lattices orders. 
This classical transition is continuous, and in the universality class of the 2D classical 
Ising model. Although the nature of the $T=0$ transition is not established at present time, 
there are suggestions that it may actually be a continuous quantum phase transition. 
The presence of the discrete sliding symmetry suggests that this is an unconventional 
quantum phase transition whose universal behavior is worth understanding.

\begin{acknowledgments}
We are grateful to Marek Biskup, Lincoln Chayes, Joel Moore,
and Jeroen van den Brink for many discussions on this problem. 
This work was supported in part by US DOE  via LDRD X1WX (Z.N.) and by the National Science Foundation 
through the grant No. DMR01-32990.
\end{acknowledgments}

\appendix

 \section{Self-duality of
``around the cube'' models in
the presence of transverse 
fields}
\label{cube}

In this appendix, we explicitly generalize
the self-duality that we obtained
earlier for plaquette models 
with a transverse field to 
cubic models with eight spin
interactions augmented by
a transverse field.
Such a duality was alluded
to in \cite{Xu03,Xu04}. 

In the below, we derive this duality by going 
back and forth from various quantum systems
to corresponding $(d+1)$ dimensional 
classical actions when different spin quantization
and spatial lattice directions are chosen.

To prove the self-duality
of such cubic systems, first 
consider the Hamiltonian
\begin{eqnarray}
H = - K \sum_{\Box \in xy \mbox{plane}} \sigma^{y} \sigma^{y} \sigma^{y} 
\sigma^{y}  \nonumber
\\ - J_{z} \sum_{\mbox{bonds along z axis}} \sigma^{z} \sigma^{z}.
\label{Ham8}
\end{eqnarray}
If we write down the classical action in a spin basis
quantized along the $\sigma^{z}$ axis, we find
\begin{eqnarray}
S_{1} = - K_{1} \sum_{\mbox{cubes in} ~ x y \tau} \sigma \sigma \sigma \sigma
\sigma \sigma \sigma \sigma  \nonumber
\\ - J_{z1} 
 \sum_{\mbox{bonds along z axis}} \sigma \sigma.
\end{eqnarray}
Here $\sinh 2 K_{1} \sinh 2 \overline{K} =1$
with $\overline{K} \equiv K \Delta \tau$, 
and $J_{z1} = J_{z} \Delta \tau$. 
Alternatively, if we write down the classical
action corresponding to Eq.(\ref{Ham8})
in a spin basis polarized along $\sigma^{y}$,
we find
\begin{eqnarray}
S_{2} = - K_{2} \sum_{\Box \in xy \mbox{plane}} \sigma \sigma \sigma \sigma 
\nonumber \\ 
- J_{z2} \sum_{\Box \in z \tau \mbox{plane}} \sigma \sigma \sigma 
\sigma.
\end{eqnarray}
Here, $K_{2} = K \Delta \tau$ and $\sinh 2 J_{z2} \sinh 2 \overline{J}_{z}= 1$
with $\overline{J}_{z} = J_{z} \Delta \tau$.
Thus, we find that the classical actions $S_{1}$ and $S_{2}$ are dual 
to each other. The classical action $S_{1}$ also corresponds
to the Hamiltonian 
\begin{eqnarray}
H_{cube} =   - K_{*} \sum_{\mbox{cubes in} ~ x y z} \sigma^{z} \sigma^{z} 
\sigma^{z} \sigma^{z}
\sigma^{z} \sigma^{z} \sigma^{z} \sigma^{z} \nonumber
\\ - h_{z*} 
 \sum_{\bf r} \sigma^{x}_{\bf r},
\end{eqnarray}
when written in a spin basis quantized along 
$\sigma^{z}$. 
Thus, $H_{cube}$ may be represented by the classical
action $S_{2}$. Putting all
of the pieces together we find that
\begin{eqnarray}
\sinh 2 K_{*} \Delta \tau \sinh 2 K_{2} = 1 \nonumber
\\ \sinh 2 h_{z*} \Delta \tau \sinh 2 J_{z2} = 1.
\end{eqnarray}

Interchanging, in the action $S_{2}$, $K \to J$,
$x \to z$, $y \to \tau$, we obtain a new action ($S_{3}$)
whose partition function is identically the same. 
By the same steps outlined above, the classical
action $S_{3}$ corresponds (via duality transformations) 
to the Hamiltonian
\begin{eqnarray}
\tilde{H}_{cube} =   - \tilde{K}_{*} \sum_{\mbox{cubes in} ~ x y z} 
\sigma^{z} \sigma^{z} 
\sigma^{z} \sigma^{z}
\sigma^{z} \sigma^{z} \sigma^{z} \sigma^{z} \nonumber
\\  - \tilde{h}_{z*} 
\sum_{\bf r} \sigma^{x}_{\bf r}.
\end{eqnarray}
This establishes the duality 
between $H_{cube}$ and $\tilde{H}_{cube}$,
\begin{eqnarray}
\tilde{K}_{*} = h_{z*} \nonumber
\\ \tilde{h}_{z*} = K_{*}.
\end{eqnarray}
Fusing these relations together, we find that
$h_{*}=K_{z*}$ constitutes a self-dual line of
$H_{cube}$.


\begin{thebibliography}{aaa}
\bibitem{AF} Some orbital interactions such as the Jahn-Teller 
interactions are of an antiferromagnetic nature $(J<0)$. However, 
in the XY model, we may identically set $J>0$
by a trivial canonical transformation
(a rotation of all spins by $\pi$ about the $\sigma_{y}$ axis- 
a flip of the spins-  on one sub-lattice). The sign of $J$
is irrelevant. Similarly, in the Hamiltonian of Eq.(\ref{an2d}),
we may set $(J_{x} \to J_{x}, J_{z} \to - J_{z})$ 
by performing a rotation by $\pi$ about the 
$\sigma^{x}$ axis on all sites belonging to one 
sub-lattice.
\end{thebibliography}
\end{document}